\documentclass[11pt,twoside]{article}

\usepackage{asp2004}
\usepackage{epsf}
\usepackage{psfig}
\usepackage{lscape}

\begin{document}
\title{Infrared Sources in the Small Magellanic Cloud: First Results}   
\author{Joshua D. Simon,\altaffilmark{1} Alberto D. Bolatto,\altaffilmark{2} 
  Sne\v{z}ana Stanimirovi\'{c},\altaffilmark{2} Ronak Y. Shah,\altaffilmark{3}
  Adam Leroy,\altaffilmark{2} and Karin Sandstrom\altaffilmark{2}} 
\affil{\altaffilmark{1}Department of Astronomy, California Institute
  of Technology, 1200 E. California Blvd., Pasadena, CA  91125 \newline
\altaffilmark{2}Department of Astronomy, University of California at Berkeley, 
601 Campbell Hall, Berkeley, CA  94720 \newline
\altaffilmark{3}Institute for Astrophysical Research, Boston University, 
725 Commonwealth Ave., Boston, MA  02215}    

\begin{abstract} 
We have imaged the entire Small Magellanic Cloud (SMC), one of the two
nearest star-forming dwarf galaxies, in all seven IRAC and MIPS bands.
The low mass and low metallicity (1/6 solar) of the SMC make it the
best local analog for primitive galaxies at high redshift.  By
studying the properties of dust and star formation in the SMC at high
resolution, we can gain understanding of similar distant galaxies that
can only be observed in much less detail. 

In this contribution, we present a preliminary analysis of the
properties of point sources detected in the \emph{Spitzer} Survey of the
Small Magellanic Cloud (S$^{3}$MC).  We find $\sim$400,000 unresolved
or marginally resolved sources in our IRAC images, and our MIPS
24~$\mu$m mosaic contains $\sim$17,000 point sources.  Source counts
decline rapidly at the longer MIPS wavelengths.  We use color-color
and color-magnitude diagrams to investigate the nature of these
objects, cross-correlate their positions with those of known sources
at other wavelengths, and show examples of how these data can be used
to identify interesting classes of objects such as carbon stars and
young stellar objects.  For additional examples of some of the
questions that can be studied with these data, please see the
accompanying contributions by Alberto Bolatto (survey information and
images), Adam Leroy (dust and gas in a low-metallicity environment),
Karin Sandstrom (far infrared-radio continuum correlation), and
Sne\v{z}ana Stanimirovi\'{c} (on a young supernova remnant in the
SMC).  The mosaic images and point source catalogs we have made have
been released to the public on our website
(http://celestial.berkeley.edu/spitzer).

\end{abstract}

%%% MAIN BODY OF TEXT GOES HERE. CONSULT "INSTRUCTIONS FOR AUTHORS USING
%%% LATEX2E MARKUP", SECTIONS 2.3-2.6 FOR HELP WITH EQUATIONS, FIGURES,
%%% AND TABLES.

\section{Introduction}   

The Magellanic Clouds are the nearest gas-rich dwarf galaxies to the
Milky Way, and are therefore excellent locations for studying star
formation.  These objects are nearby, so that individual stars and
H~{\sc ii} regions can be studied in detail, and their low
metallicities provide a different environment in which star formation
theories developed for our Galaxy can be tested.  For the Small
Magellanic Cloud (SMC) in particular, the metallicity begins to
approach the value expected for galaxies at high redshift, making the
SMC a very useful prototype for understanding star formation in the
distant universe.

In this contribution, we present some initial results from the
\emph{Spitzer} Survey of the Small Magellanic Cloud (S$^{3}$MC),
including infrared color-magnitude diagrams of the entire SMC, and a
search for young stellar objects (YSOs) in the SMC H~{\sc ii} region
  N66.

\section{Observations and Data Reduction}

The S$^{3}$MC is a project to map the SMC with \emph{Spitzer} in all
seven Infrared Array Camera \citep[IRAC;][]{irac} and Multiband
Imaging Photometer for \emph{Spitzer} \citep[MIPS;][]{mips} bands.
The data cover an area of $\sim2.5$~deg$^{2}$, including the entire
bar and wing of the SMC.  We constructed mosaic images from the
individual Basic Calibrated Data frames using the Mosaicking and Point
Source Extraction (MOPEX) software provided by the \emph{Spitzer}
Science Center.  Further details of the observations and the data
processing will be described by Bolatto et al. (in preparation).  We
performed point source photometry on the mosaic images with the
Astronomical Point Source Extraction (APEX) tasks in the MOPEX package
\citep{apex}.  The resulting photometry tables have been made freely
available on our project website, and will be updated in the future as
we produce improved data products.

\section{Results}
\label{results}

We detected a total of over 406,000 point sources in the S$^{3}$MC
mosaics.  The images in the two most sensitive bands (IRAC 3.6~$\mu$m
and 4.5~$\mu$m) contain approximately 290,000 point sources each,
while the 5.8~$\mu$m mosaic contains 80,000 and the 8.0~$\mu$m mosaic
62,000.  Because of the reduced sensitivity and the declining spectral
energy distributions for most types of objects, the source counts in
the MIPS images are much lower, with $\sim17,000$ objects detected at
24~$\mu$m and 1700 detected at 70~$\mu$m.  We did not attempt point
source photometry at 160~$\mu$m because there did not appear to be any
point sources present.  

\begin{figure}[!t]
%\epsscale{1.2}
\plottwo{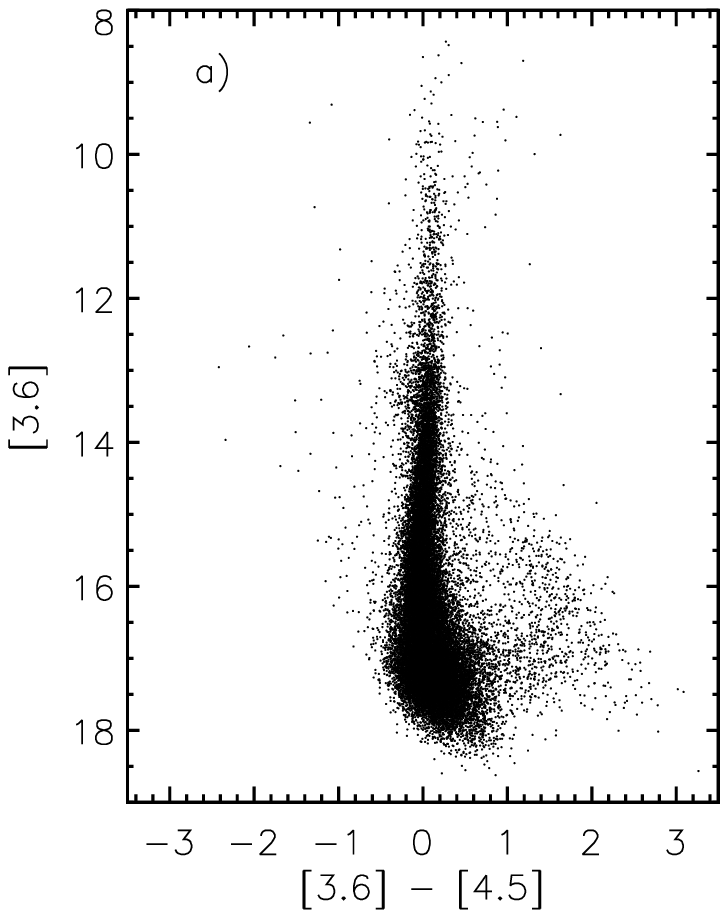}{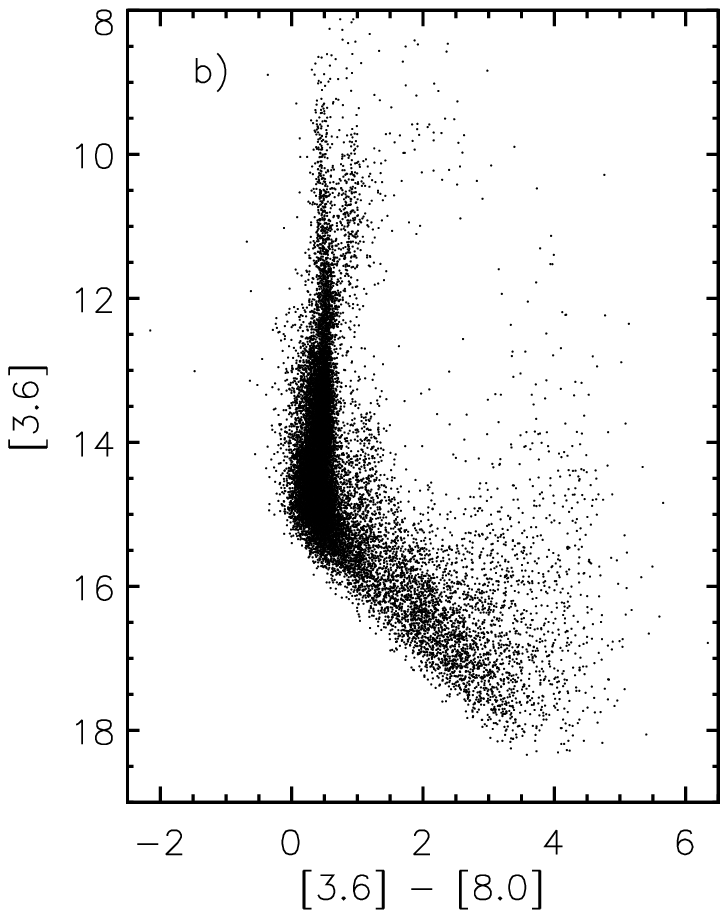}
\caption{(\emph{a}) IRAC [3.6] vs. [3.6]~$ - $~[4.5] color-magnitude
  diagram of the 201,000 stars that are detected in both bands.  The
  stars are concentrated in a vertical plume at a color of zero
  because these wavelengths are on the Rayleigh-Jeans tail of the
  spectral energy distribution for most types of stars.  To save space
  on astro-ph, only every sixth point is plotted.  (\emph{b}) IRAC
  [3.6] vs. [3.6]~$ - $~[8.0] color-magnitude diagram of the 40,000
  stars that are detected in both bands.  The longer wavelength
  baseline provided by the 8.0~$\mu$m data allows a much better
  separation of normal stars (still in the vertical plume near zero
  color) and YSOs (the population of very red sources on the right
  side of the diagram), and also splits carbon stars off from the main
  distribution. To save space on astro-ph, only every other point is
  plotted.}
\label{cmd1}
\end{figure}

In Figure \ref{cmd1} we display the color-magnitude diagrams (CMDs)
for the entire SMC in IRAC bands 1 and 2 and bands 1 and 4 (note that
we use Vega magnitudes).  The CMD for bands 1 and 2 emphasizes
the massive main sequence stars, red giants, and asymptotic giant
branch (AGB) stars, which are concentrated in a narrow vertical
sequence very close to a [3.6]~$ - $~[4.5] color of zero.  There is
also a scattering of red objects visible to the right of the main
plume, which likely represents stars that have some associated dust
emission.  Many of the fainter objects are probably protostars still
surrounded by disks, and the brighter stars are expected to be AGB
stars of various types. The 3.6~$\mu$m luminosity function shows a
significant break at [3.6]~$\approx $~12.7 that may correspond to the
luminosity of the tip of the red giant branch.  In the CMD for bands 1
and 4 in Figure \ref{cmd1}\emph{b} the separation of dusty objects and
normal stars is much improved by the longer wavelength baseline.  At
very red colors a sample of protostars can now be cleanly selected,
and carbon stars have been split off to the right of the main AGB
sequence.  These data will thus be very useful for selecting samples
of dusty stars that might be missed or confused with other types of
objects at optical wavelengths.

\subsection{Young Stellar Objects in N66}

\begin{figure}
\plottwo{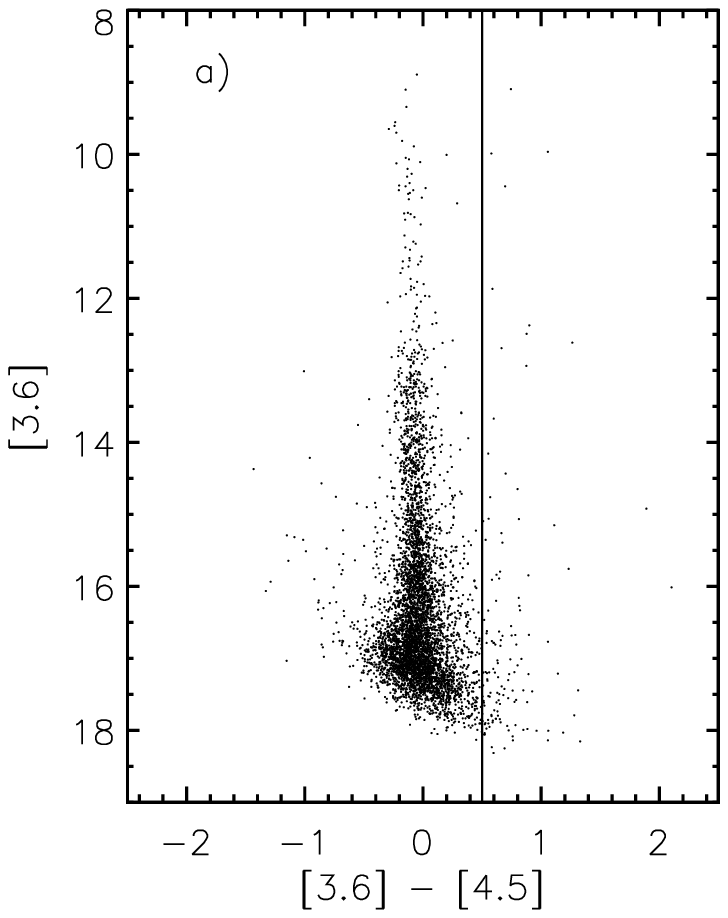}{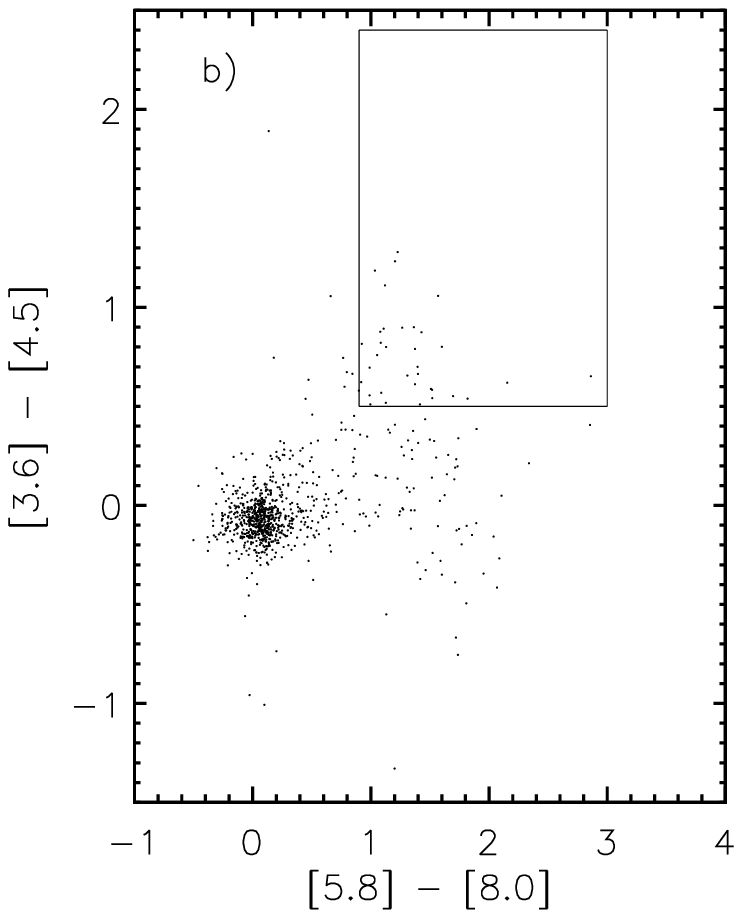}
\caption{(\emph{a}) IRAC [3.6] vs. [3.6]$ - $[4.5] color-magnitude
  diagram of N66 (compare to figure \ref{cmd1}\emph{a}).  The vertical
  line drawn at a color of [3.6]$ - $[4.5]$ = $ 0.5 represents the
  reddest color expected for normal, unreddened stars.  Sources to the
  right of this line are considered candidate young stellar
  objects. (\emph{b}) IRAC color-color diagram of N66.  The box in the
  upper right outlines the colors characteristic of the YSO models of
  \citet{whitney04b}.}
\label{n66cmd}
\end{figure}

As we showed in \S 3, it is straightforward to use the
IRAC photometry to select YSOs (see Figure \ref{cmd1}\emph{b}).  As a
case study, we have examined the point source population of the most
massive star-forming region in the SMC, N66.  In Figure
\ref{n66cmd}\emph{a} we display a CMD for N66.  The 136 stars with
[3.6]~$ - $~[4.5]~$ > 0.5$ are candidate YSOs.  In Figure
\ref{n66cmd}\emph{b} we plot a color-color diagram of every star
in N66 that is detected in all four IRAC bands, and outline the
region of color-color space in which the YSO models of
\citet{whitney04b} are located.  There are 34 stars that meet this
more stringent criterion and are almost certainly young stellar
objects, representing the largest sample of early-class protostars
yet identified in the SMC.  In a future paper, we will carry out
detailed modeling of the spectral energy distributions of the YSOs
in N66 so that we can classify each of the stars and study whether
their properties differ from YSOs in the Milky Way (Simon et al., 
in preparation).

\acknowledgements 

JDS gratefully acknowledges the support of a Millikan Fellowship
provided by the California Institute of Technology.  We thank David
Makovoz for his extensive assistance with MOPEX.  This work is based
on observations made with the \emph{Spitzer Space Telescope}, which is
operated by the Jet Propulsion Laboratory, California Institute of
Technology, under a contract with NASA.  Support for this work was
provided by NASA through an award issued by JPL/Caltech.

%%% THE BIBLIOGRAPHY
%%%
%%% CONSULT SECTION 3 OF "INSTRUCTIONS FOR AUTHORS" FOR HOW TO USE NATBIB.
%%% AUTHORS ARE ENCOURAGED TO USE EITHER THE "THEBIBLIOGRAPY" ENVIRONMENT
%%% BY UNCOMMENTING (DELETING THE "%" SYMBOL) THE COMMANDS BELOW, OR BY
%%% USING THE BIBTEX ENVIRONMENT. TO FIND OUT WHICH IS APPLICABLE TO YOUR
%%% CONTRIBUTION, CONSULT THE VOLUME EDITORS FOR YOUR PROCEEDINGS.
%%%

\end{document}